# Observation of a Helical Luttinger-Liquid in InAs/GaSb Quantum Spin Hall Edges

Tingxin Li[1,4], Pengjie Wang[1,4], Hailong Fu[1,4], Lingjie Du[2], Kate A. Schreiber[3], Xiaoyang Mu[1,4], Xiaoxue Liu[1,4], Gerard Sullivan[5], Gábor A. Csáthy[3], Xi Lin[1,4], Rui-Rui Du[1,2,4]*

[1]*International Center for Quantum Materials, School of Physics, Peking University, Beijing 100871, China*

[2]*Department of Physics and Astronomy, Rice University, Houston, Texas 77251-1892, USA*

[3]*Department of Physics and Astronomy, Purdue University, West Lafayette, Indiana 47907, USA*

[4]*Collaborative Innovation Center of Quantum Matter, Beijing 100871, China*

[5]*Teledyne Scientific and Imaging, Thousand Oaks, California 91603, USA*

*Abstract*

We report on the observation of a helical Luttinger-liquid in the edge of InAs/GaSb quantum spin Hall insulator, which shows characteristic suppression of conductance at low temperature and low bias voltage. Moreover, the conductance shows power-law behavior as a function of temperature and bias voltage. The results underscore the strong electron-electron interaction effect in transport of InAs/GaSb edge states. Because of the fact that the Fermi velocity of the edge modes is controlled by gates, the Luttinger parameter can be fine tuned. Realization of a tunable Luttinger-liquid offers a one-dimensional model system for future studies of predicted correlation effects.

It is well known that electron-electron interactions play a more important role in one-dimensional (1D) electronic system than that in higher dimensional systems. In 1D system, interactions cause electrons to behave in a strongly correlated way, so under very general circumstances, 1D electron systems can be described by Tomonaga-Luttinger liquid (LL) theory [1,2] instead of mean-field Fermi liquid theory. A Luttinger parameter $K$ characterizes the sign and the strength of the interactions: $K < 1$ for repulsion, $K > 1$ for attraction, and $K = 1$ for non-interacting case. Confirmations of LL have been examined in various materials, such as carbon nanotubes [3-5], semiconductor nanowires [6], cleaved-edge-overgrowth 1D channel [7], as well as fractional quantum Hall (FQH) edge states [8], respectively for spinful or chiral Luttinger-liquids. The experimental hallmarks of LL are a strongly suppressed tunneling conductance and a power-law dependence of the tunneling conductance on temperature and bias voltage [3-5,8]. In a weakly disordered spinful LL, transport experiments showed that the conductance reduces from the quantized value as the temperature is being decreased [6,7].

The quantum spin Hall insulator (QSHI), also known as two-dimensional (2D) topological insulator (TI), is a topological state of matter supporting the helical edge states, which are counter-propagating, spin-momentum locked 1D modes protected by time reversal symmetry. It has recently attracted a lot of interest due to their peculiar helical edge properties and potential applications for quantum computation [9-18]. Experimentally, QSHI has been realized in HgTe quantum wells (QWs) [14] and in InAs/GaSb QWs [16-18]. In both cases, quantized conductance plateaus have been observed in devices with edge length of several micrometers [14,18], implying ballistic transport in the edges. On the other hand, devices with longer edges have lower values of conductance [14,17,18], indicating certain backscattering processes occurred inside helical edges. In principle, single-particle elastic backscattering is forbidden in helical edges due to the protection of time reversal symmetry. Therefore, inelastic and/or multiparticle scattering should be the dominating scattering mechanisms, which would lead to temperature-dependent edge conductivity [19-25]. However, in InAs/GaSb QSHI, existing experiments surprisingly show that the edge conductance is independent of temperature from 20 mK up to 30 K for both small and large samples [17,18].

The (spinless) helical LL behavior is here observed in the helical edges of InAs/GaSb QWs where the Fermi velocity of edge states is low (in the order of $v_F$ ~$10^4$ m/s), resulting in strong interaction effects. Fig. 1a shows the schematic drawing of spinful LL, chiral LL, and helical LL. The dispersion of a spinful LL is linearized around the Fermi level, in comparison to the non-interaction case. The left and right moving branches of a spinful LL are always separated by a momentum of roughly $2k_F$. As for the helical LL, two branches cross at the Dirac point, thus a unique momentum-conserving umklapp scattering process [23,24] could occur near the Dirac point, in a generic ($S_z$ symmetry broken) helical LL with sufficiently strong interactions. Also the degrees of freedom in a helical LL are only half as in a spinful LL. Fig. 1b schematically depicts the electron transport in a helical LL, where counter-propagating, strongly correlated electrons have soliton-like excitations in the ballistic transport regime.

The wafer structures for experiments are shown in Fig. 2a. Experiments are performed in two millikelvin dilution refrigerators (DR) instrumented for fractional quantum Hall effect studies, one of them having attained ~7 mK electron temperature by using a He-3 immersion cell [26], as depicted in Fig. 2b. The second DR has attained about 30 mK electron temperature [27]. The quantity, $T$, mentioned in the following text refer to electron temperature. Devices investigated are made with a Schottky-type front gate, showing less hysteresis effect than previous devices [17,18]. In these experiments, care is exercised to exclude spurious effects such as those from nonlinear contacts, or leaking conductance through bulk states, etc. (see section IV and VI of Supplemental Material [28]).

Fig. 2c shows the four-terminal longitudinal resistance $R_{xx}$ as a function of the front gate voltage $V_{front}$ in a 20×10 μm² six-terminal Hall bar device (wafer A) biased with different excitation currents at $T$~6.8 mK. $R_{xx}$ was measured using standard low frequency (17 Hz) lock-in techniques. As the Fermi level is tuned into the QSHI gap via front gate, the $R_{xx}$ shows a peak. Remarkably, peak values decrease with increasing current $I$, which indicates the helical edge has nonlinear conductance characteristics. Fluctuations can be observed in the $R_{xx}$ peak region, and the amplitude of the fluctuations decreases with the increasing of $I$ or $T$. Moreover, these fluctuations have an amplitude larger than the background noise level, and to some extent they are reproducible (see section III of Supplemental Material [28]). The inset of Fig. 2c shows the helical edge conductance $\bar{G}_{xx}$ (conductance of the averaged $R_{xx}$ peaks) as a function of $T$. It can be seen that for each $I$ value, there exists a $T$-independent range for $\bar{G}_{xx}$. However, the lower the current is, the narrower the $T$-independent range. The most likely explanation is that the helical edge conductance does not show $T$-dependence for the $eV \gg k_BT$ regime, where $k_B$ is the Boltzman constant. Notice that previous experiments [17,18] all used relatively high $I$, leading to a large e$V$ across the helical edge, so the measured edge conductance were found to be $T$-independent in a large range.

We note that all devices measured here have shown these characteristic nonlinear transport. In the following we will focus on the systematic results measured from a mesoscopic two-terminal device (wafer B, edge length ~ 1.2 μm). $R_{xx}$ was measured in a quasi-four-terminal configuration, and a series resistance ~1.9 kΩ has been subtracted for all data points. Fig. 3a shows several $R_{xx}$-$V_{\text{front}}$ traces taken at different temperatures with a large bias current (500 nA). The quantized resistance plateau of $h/2e^2$ persists from 30 mK to 2 K, conforming to the behavior for $eV \gg k_BT$; eventually the total conductance increases at higher $T$ ($T > 2$ K) due to the delocalization of bulk states (inset in Fig. 3a). Fig. 3b shows the $T$-dependence of $\bar{G}_{xx}$ with two different currents from 30 mK to 1.2 K, where the bulk conductance is negligible. The measured $\bar{G}_{xx}$ with 0.1 nA excitation current can be fitted with a power-law function of $T$, $\bar{G}_{xx} \propto T^\alpha$ with exponent $\alpha \approx 0.32$. As for the $I = 2$ nA case, $\bar{G}_{xx}$ is independent of $T$ in the regime where $eV \gg k_BT$, then following the same power-law as the $I = 0.1$ nA case at higher $T$ ($T > 500$ mK).

A reasonable explanation for these striking experimental observations should be based on the strong electron-electron interactions in the helical edge states of InAs/GaSb. Note that helical edge states have a topological stability that is insensitive to nonmagnetic disorder and weak interactions [11-13,19], which is in contrast with spinful LL where the conductance vanishes at $T = 0$ even for an arbitrarily weak disorder and interaction [2,29,30]. However, in the strong interaction regime ($K < 1/4$), correlated two-particle backscattering (2PB) processes are relevant [12,13,19-21] in helical edge even with a single trivial impurity (here they could be charge puddles [19,25], defects of crystalline, Rashba spin-orbit coupling [21,22], and so on), breaking the 1D helical edge into segments, thus forming a "Luttinger-liquid insulator" at $T = 0$. At low but finite $T$, $\bar{G}_{xx}$ is restored by tunneling [12,19] of excitations with fractional charge $e/2$ between energy minima inside helical edges, resulting in $\bar{G}_{xx}(T) \propto T^{2(1/4K-1)}$. A breakdown of such tunneling processes takes place when the external energy (temperature or bias voltage) is larger than the energy of the potential pinning the edge states. Therefore, the quantized conductance plateau for QSHI is recovered at large bias voltage, as we have observed.

$K$ value of a helical LL can be estimated by formulas given in Ref. [19,31] (see section V of Supplemental Material [28]). $K$ in HgTe QWs is about 0.8 (Ref. [31]), indicating a weak interaction regime. In InAs/GaSb QWs, $K \sim 0.22$ for wafer B, is in the strong interaction regime. From the power-law exponent obtained from experiments, we deduce $K \sim 0.21$, which is in good agreement with theoretical estimations.

Bias voltage dependence has also been systematically measured for the same 1.2 μm device. The inset in Fig. 4 shows the measured edge *differential* conductance $dI/dV$ as a function of $V_{dc}$ (the applied dc bias voltage) at various temperatures, on a double logarithmic scale. At low bias $eV_{dc} \ll k_BT$, $dI/dV$ is constant with $V_{dc}$ but the value depends on $T$. At higher bias, $dI/dV$ increases with $V_{dc}$ follows an approximate power-law, and the fitted exponent is about 0.37. Further increasing $V_{dc}$, $dI/dV$ begins to deviate from the power-law behavior, tending to saturate toward the quantized value of $2e^2/h$. Furthermore, all the data points except the saturation region collapse onto a single curve if the differential conductance is scaled by $T^\alpha$ and plotted versus $eV_{dc}/k_BT$, as shown in Fig. 4. Similar scaling relations have been observed previously in spinful LL [3-5] and chiral LL [8], and were taken as a critical evidence of LL. Here the observed scaling relation could be suggestive for the internal tunneling processes mentioned above [12,19], since there is not any man-made tunneling barrier in our devices.

The preceding analyses are based on single impurity case, but they should still be valid for multiple, *isolated* impurities. Randomly distributed impurities may introduce a series of tunneling barriers into the helical edge, making the edge more resistive, but would not break the power-law relations. On the other hand, even without explicit impurities, uniform 2PB (umklapp) term can arise in the presence of anisotropic spin interactions [12] or just in a $S_z$ symmetry broken helical liquid as mentioned in Ref. [23,24]. Such umklapp term in combination with strong electron-electron interaction ($K < 1/2$) leads to gap opening in the helical edge [12,32,33], or to the formation of a 1D Wigner crystal phase [34] at ultralow temperatures. When increasing the temperature or bias voltage, the umklapp processes become weakened and non-uniform so the gap becomes 'soft', resulting in a finite conductance [35]. Future experiments such as quantum point contact [31,36] and shot-noise [19,20] measurements could in principle reveal the microscopic physical processes inside such strongly interacting helical edge states.

In conclusion, in InAs/GaSb QSHI we observe a strong suppression of the helical edge conductance at low temperature and bias voltage, which suggests that strong electron-electron interactions in the helical edges should lead to a correlated electronic insulator phase at $T = 0$ and vanishing bias voltage. Due to the fact that the bulk gaps (hence the $v_F$ of edge states) in InAs/GaSb materials can be engineered by molecular-beam epitaxy growth and gating architectures, the electron-electron interactions can be fine-tuned, leading to a well-controlled model system for studies of 1D electronic and spin correlation physics. It's well known that [9,10] the QSHI helical edge states coupled with superconductors can support Majorana zero modes. More interestingly, the presence of strong interactions promotes these Majorana modes splitting into $Z_4$ parafermionic modes [32,33], which are promising for universal, decoherence-free quantum computation. The Josephson junction mediated by interacting QSHI edge states creates a pair of parafermions, yield a novel $8\pi$-Josephson effect reflecting the tunneling processes of $e/2$ charge quasiparticles between superconductors. Further studies of interaction effects on the helical edge states in InAs/GaSb system would be necessary to advance in this direction.

We acknowledge discussions with C. L. Kane, L. I. Glazman, and C. J. Wu. The work at PKU was supported by NBRPC Grant (No. 2012CB921301 and No. 2014CB920901). R.R.D. was supported by NSF Grant (No. DMR-1207562 and DMR-1508644), L.J.D. was supported by DOE Grant (No. DE-FG02-06ER46274). The ultra-low temperature transport measurements at Purdue were supported by DOE Grant (No. DE-SC0006671).

* Email: rrd@rice.edu

**Figure Captions**

FIG. 1 Family of Luttinger-liquids. (a) Schematic drawing of energy dispersions for spinful LL, chiral LL, and helical LL, respectively; for the spinful LL, two straight lines illustrate the linearized dispersion, corresponding to the left and right moving branches, respectively. In the chiral LL, strongly correlated, spin degenerated electrons move in only one direction. As for the helical LL, the left and right moving branches cross at the Dirac point, and electrons with opposite spins move in opposite directions. (b) Schematic drawing of the electron transport in a helical LL.

FIG. 2 (a) Specific structures of two InAs/GaSb wafers used for experiments. (b) Schematic drawing of the He-3 immersion cell [26]. Orange, light grey, dark grey and black parts represent copper, polycarbonate, silver, and the sample, respectively. The cell is attached to the mixing chamber of the DR and filled with liquid He-3 through a capillary. Contacts of the sample are soldered with indium to several heatsinks which are made of 100-500 nm silver powder sintered on to silver wires. (c) $R_{xx}$ of a 20×10 μm$^2$ Hall bar made by wafer A versus $V_{front}$ at $T$~6.8 mK biased with different currents. Inset in **c**, helical edge conductance $\bar{G}_{xx}$ as a function of $T$. At 0.1 nA, $\bar{G}_{xx}$ begins to change for $T > 60$ mK, and the critical $T$ is about 160 mK for the 1 nA case. As for the 10 nA case, there is no obvious change of $\bar{G}_{xx}$ below 250 mK.

FIG 3. Temperature dependence for a mesoscopic device (wafer B, edge length ~1.2 μm). (a) $R_{xx}$−$V_{front}$ traces taken at 30 mK, 350 mK, 1 K, and 2 K with 500 nA excitation current. Quantized resistance plateau of $h/2e^2$ persists from 30 mK to 2 K. Inset in (a), plateau conductance increases at higher temperature due to delocalized bulk carriers. (b) Temperature dependence of the helical edge conductance $\bar{G}_{xx}$ with $I = 0.1$ nA, and 2 nA. The straight line on the log-log plot indicates a power-law behavior $\bar{G}_{xx} \propto T^{0.32}$. Inset in (b) shows the SEM image of the device.

FIG 4. Bias voltage dependence for a mesoscopic device (wafer B, edge length ~1.2 μm). The inset shows $V_{dc}$ dependence of the edge differential conductance d$I$/d$V$ measured at $T = 50$ mK, 100 mK, 350 mK, and 1 K, with the ac modulation current $I_{ac} = 0.1$ nA. The solid line indicates a power-law of $dI/dV \propto V_{dc}^{0.37}$. The main plot illustrates all the measured data

points except the saturation region collapse onto a single curve by scaling the measured d*I*/d*V*.

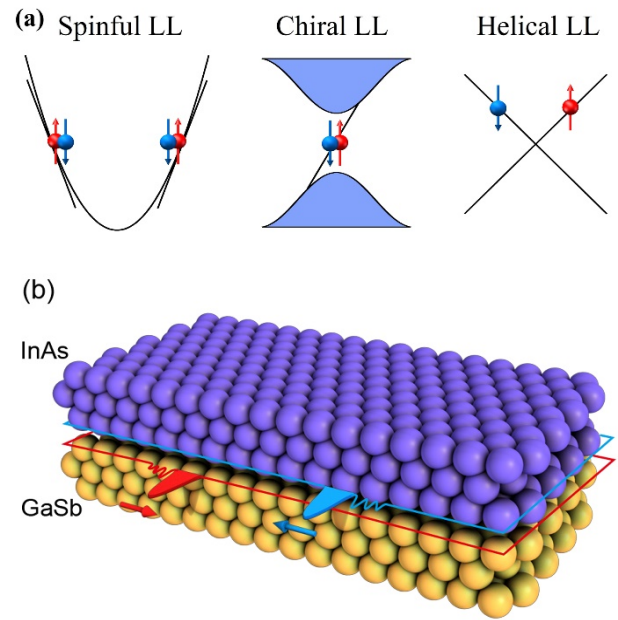

**Figure 1**

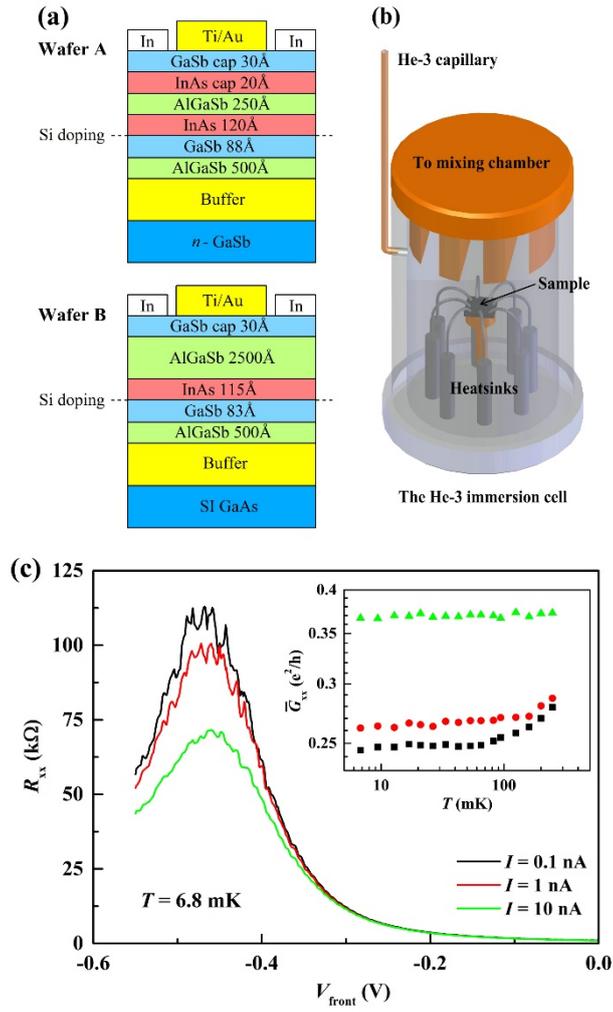

**Figure 2**

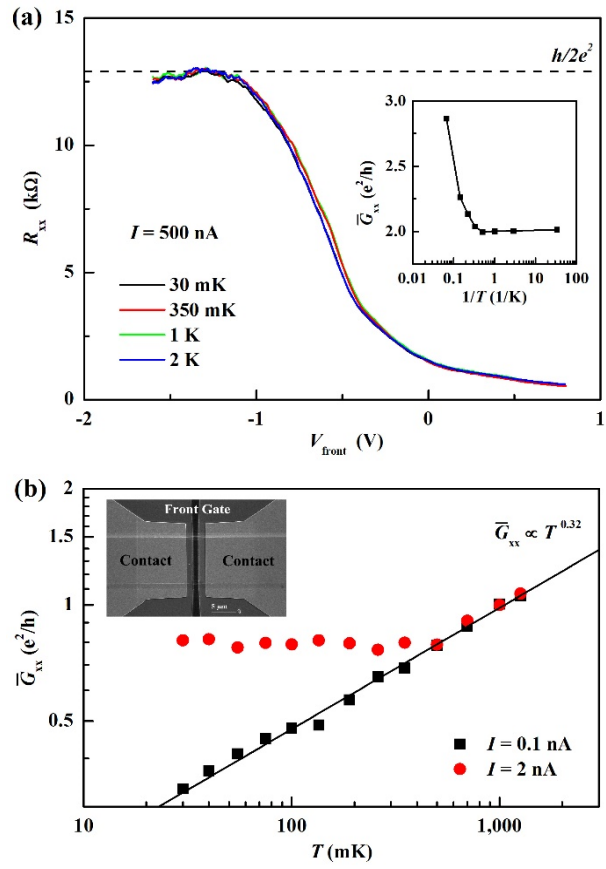

**Figure 3**

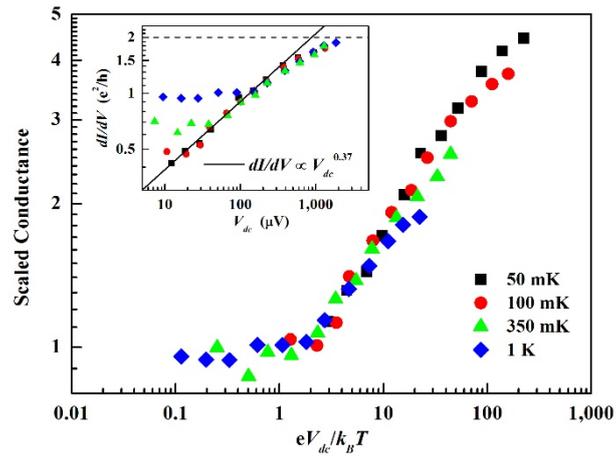

**Figure 4**

# Supplemental Materials:

# Observation of a Helical Luttinger-Liquid in InAs/GaSb Quantum Spin Hall Edges

Tingxin Li[1,4], Pengjie Wang[1,4], Hailong Fu[1,4], Lingjie Du[2], Kate A. Schreiber[3], Xiaoyang Mu[1,4], Xiaoxue Liu[1,4], Gerard Sullivan[5], Gábor A. Csáthy[3], Xi Lin[1,4], Rui-Rui Du[1,2,4]*

[1]*International Center for Quantum Materials, School of Physics, Peking University, Beijing 100871, China*

[2]*Department of Physics and Astronomy, Rice University, Houston, Texas 77251-1892, USA*

[3]*Department of Physics and Astronomy, Purdue University, West Lafayette, Indiana 47907, USA*

[4]*Collaborative Innovation Center of Quantum Matter, Beijing 100871, China*

[5]*Teledyne Scientific and Imaging, Thousand Oaks, California 91603, USA*

**I Details of wafers and devices**

The semiconductor wafers of InAs/GaSb QWs were grown by molecular beam epitaxy (MBE) technique. Both wafers are silicon-doped for getting an insulating bulk [S1,S2]. The density $n$ and mobility $\mu$ can be deduced from the magneto-transport data. The mobility of wafer A is about 64,000 cm$^2$/Vs at a density of about $6.5\times10^{11}$ cm$^{-2}$, and the mobility of wafer B is about 13,000 cm$^2$/Vs at a density of about $6.4\times10^{11}$ cm$^{-2}$. Fig. S1 shows the $R_{xx}$–$V_{front}$ traces and $n$–$V_{front}$ traces of typical devices for both wafers. By fitting the measured data points, the inverted band crossing density $n_{cross}$ can be deduced, which are ~$1.3\times10^{11}$ cm$^{-2}$ and ~$0.6\times10^{11}$ cm$^{-2}$ for wafer A and wafer B, respectively.

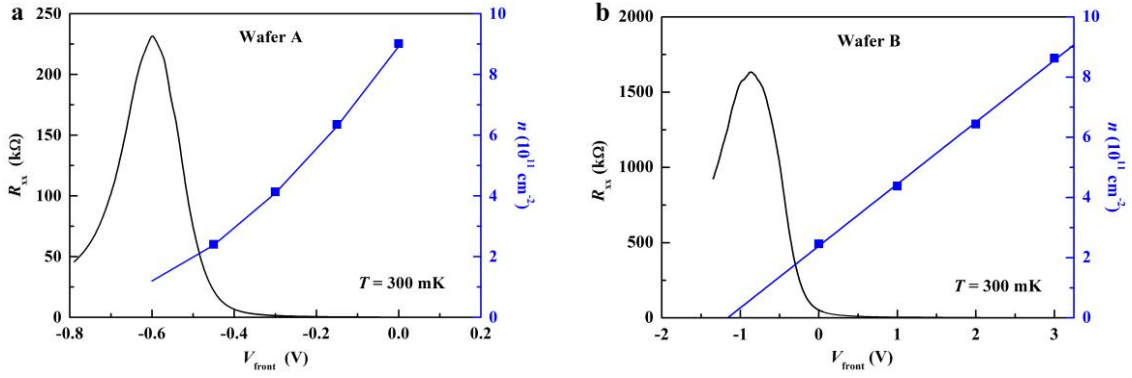

**Figure S1 a** and **b** show the $R_{xx}$–$V_{front}$, and $n$–$V_{front}$ traces of typical six-terminal Hall bar devices (edge length ~60 μm) for wafer A and wafer B, respectively. Magneto-resistance measurements were performed at 300 mK. For wafer A, the relation between $n$ and $V_{front}$ is not linear, since the front gate is very close to the QWs.

Device processing consisted of the following steps. Mesas were defined by wet etching. For multiple-terminal devices, contacts were made by directly soldering indium at 300 ℃. The contacts of two-terminal devices consisted of germanium (Ge), palladium (Pd), and gold (Au) layers, deposited by E-beam evaporation, then annealed at 250℃. A 100 nm layer of aluminum or 10 nm/90 nm layers of titanium/gold were deposited as front gate. Optical microscope image of a 20×10 μm² six-terminal Hall bar device is depicted in Fig. S2, and the inset of figure 3b in the main text shows the SEM image of a mesoscopic two-terminal device.

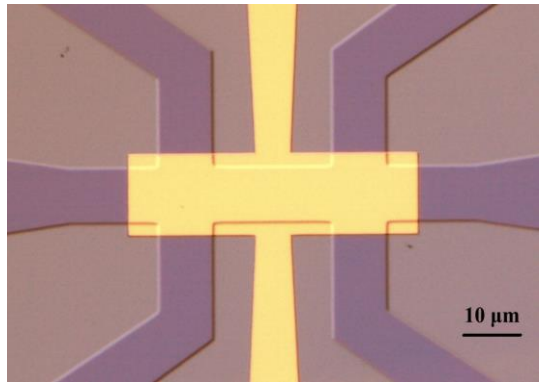

**Figure S2** The optical microscope image of a 20×10 μm² six-terminal Hall bar device made by wafer A.

The front gate and contacts must not overlap to prevent electric short. Therefore, a series resistance needs to be subtracted for the two-terminal devices. We regard the resistance value at very positive $V_{front}$ as the series resistance, because in this case the resistance of sample under the front gate is very small.

**II Device statistics**

Total 62 devices were fabricated in 8 batches, among those 43 were tested at room temperature and 4 K, only 29 exhibited reliable gating characteristics were further tested at 300 mK or lower temperature. Of these, 22 devices were measured with different bias voltage, and all of them show nonlinear characteristic of the helical edge conductance for $eV \gg k_BT$ regime. 4 mesoscopic devices and 4 large devices of high quality were systematically measured at DR for the temperature dependence and the bias voltage dependence.

**III Gate hysteresis and $R_{xx}$ fluctuations**

Fig. S3a illustrates four $R_{xx}$-$V_{front}$ traces of the 20×10 μm² Hall bar mentioned in the main text. It can be seen that the front gate shows a bit of hysteresis effect, since the sample is always a little more resistive on downward sweeps ($V_{front}$ from 0 V to -0.55 V). Such hysteresis used to be very large in devices using SiN or oxide layers as gate dielectric [S1]. When sweep front gate in the same direction, for instance traces a, c or traces b, d in Fig. S3a, the data are quite consistent, which make sure the change of $R_{xx}$ peak values is not due to the gate hysteresis.

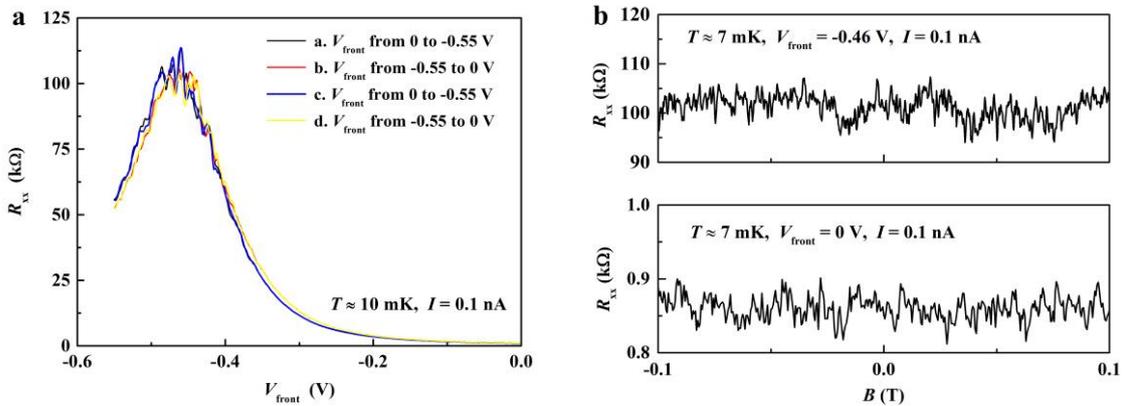

**Figure S3 a** shows four $R_{xx}$-$V_{front}$ traces at ~10 mK with 0.1 nA excitation current (17 Hz). **b** shows $R_{xx}$ fluctuations under small perpendicular magnetic field with different front gate voltages.

From Fig. S3a, it can be also seen that traces are smooth when sample is in the electron regime, and fluctuations emerge at the $R_{xx}$ peak region, corresponding to the sample being tuned into QSHI phase. The fluctuations are reproducible to some extent. Fig. S3b shows the $R_{xx}$-$B$ curves measured at ~7 mK with 0.1 nA excitation current (17 Hz), with $V_{front}$ = 0 V, and $V_{front}$ = -0.46 V, respectively. The sweep rate of magnetic field is about 0.02 T/min. For the $V_{front}$ = 0 V case, fluctuations are within 80 ohms, which can be considered as the background noise. For $V_{front}$ = -0.46 V, the amplitude of fluctuations is more than 5,000 ohms, which is much larger than the background noise level. Moreover, there is no obvious change of the fluctuations under small perpendicular magnetic field. Also note that small perpendicular field cannot break the QSHI state in InAs/GaSb, consisting with the previous results [S1-S2].

**IV Corbino disks**

In Corbino disks, edge transport is shunted via concentric contacts, thus conductance measurements probe bulk properties exclusively. The outer diameter and the inner diameter of Corbino disks for measurements are 1.2 mm and 0.6 mm, respectively. The contacts are made by annealed Ge/Pd/Au, and front gates are also Schottky-type. For wafer A, the bulk resistance per square is ~88 MΩ at 30 mK, and ~1 MΩ at 300 mK. From the Arrhenius plot (Fig. S4a), the energy gaps can be deduced by fitting $G_{xx} \propto \exp(-\Delta/2k_B T)$, where $\Delta$ is the energy required to create a pair of electron-hole over the gap. The hybridization induced minigap and the silicon-doping induced localization gap for wafer A are ~28 K and ~1.3 K, respectively.

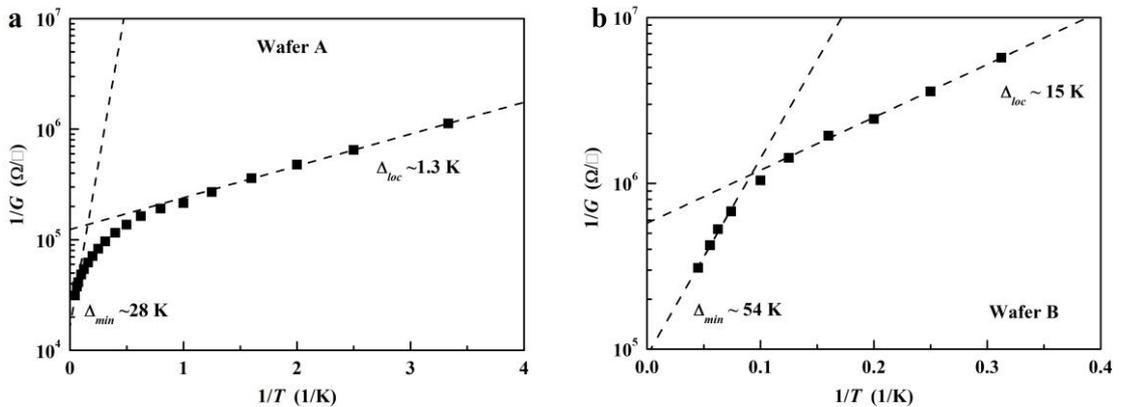

**Figure S4 a** Arrhenius plot of a Corbino disk made by wafer A. **b** Arrhenius plot of a Corbino disk made by wafer B. The bias voltage during the measurements is 1 mV.

Even at 300 mK, the bulk resistance per square of wafer B is >>100 MΩ (Fig. S5), which indicates wafer B has a truly insulating bulk, in comparison to wafer A. The minigap and the localization gap for wafer B are ~54 K and ~15 K, respectively (Fig. S4b). At 300 mK, the bulk states of wafer B show bias independent behavior up to 50 mV (Fig. S5). On the basis of the above results, we draw a conclusion that the observed $T$ dependent and bias voltage dependent characteristics of conductance are from helical edge states instead of bulk states.

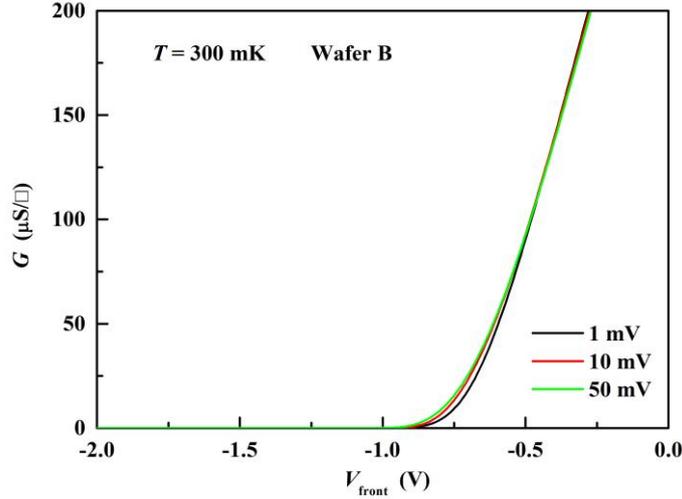

**Figure S5** Bias voltage dependence measurements of a Corbino disk made by wafer B at $T = 300$ mK. Even for the 50 mV case, the bulk resistance per square of wafer B is still >>100 MΩ.

## V Estimation of Luttinger parameter $K$

$K$ in a QSHI can be estimated by [S3-S4]

$$K = \left[1 + \frac{2}{\pi^2} \frac{e^2}{\varepsilon \hbar v_F} \ln\left(\frac{d}{\max\ \xi, w}\right)\right]^{-1/2}$$

where $\varepsilon$ is the bulk dielectric constant; $d$ is the distance from the QWs layers to a nearby metallic gate acts as a screening length for Coulomb potential; $w$ is the thickness of the QWs; assuming a linearly dispersing helical edge state, hence $v_F = \frac{1}{\hbar}\frac{\partial E}{\partial k} \sim \frac{E_{gap}}{2\hbar k_{cross}}$, where $v_F$ is the Fermi velocity of the helical edge state, $E_{gap}$ is the energy gap of the bulk QSHI, and $k_{cross} = \sqrt{2\pi n_{cross}}$; $\xi = 2\hbar v_F / E_{gap}$ is the evanescent decay length of the edge state wave function into the bulk QSHI.

For HgTe QWs [S3-S4],

$\varepsilon \approx 15$, $v_F \approx 5.5 \times 10^5$ m/s, $\xi \approx 30$ nm, $d \approx 150$ nm, $w \approx 12$ nm, so $K \approx 0.8$;

For InAs/GaSb QWs wafer A,

$\varepsilon \approx 12.5$, $v_F \approx 2.0 \times 10^4$ m/s, $\xi \approx 9$ nm, $d \approx 40$ nm, $w \approx 20$ nm, so $K \approx 0.24$;

For InAs/GaSb QWs wafer B,

$\varepsilon \approx 12.5$, $v_F \approx 5.7 \times 10^4$ m/s, $\xi \approx 16$ nm, $d \approx 260$ nm, $w \approx 20$ nm, so $K \approx 0.22$.

**VI Contact resistance**

In order to measure the contact resistance accurately, we made a six-terminal Hall bar device (as shown in Fig. S6) without front gate by wafer A. Contacts are made by directly soldering Indium. The Hall bar is designed with clear aspect ratio so that we can also calculate the resistance of arms. The length of 'a' to 'e' in Fig. S6 is 200 μm, 400 μm, 150 μm, 100 μm, and 300 μm, respectively. The lock-in measured (17 Hz, 1 nA) four-terminal resistance $R_{\text{4-terminal}}$ (*I*: A-B, *V*: C-D) at 30 mK is about 1008 ohms, and the three-terminal resistance $R_{\text{3-terminal}}$ (*I*: A-B, *V*: D-B) at 30 mK is about 884 ohms, so the contact resistance of contact B is about 212 ohms. We have measured all six contacts and result in close values (~200 ohms). For Ge/Pd/Au type contacts, the contact resistances are similar.

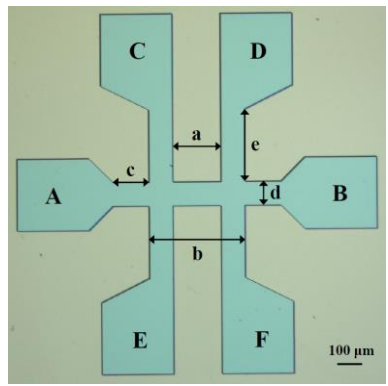

**Figure S6** Optical microscope image of a standard six-terminal Hall bar for measuring contact resistance.

For non-ohmic contacts, there is always a capacitance component in parallel to the resistance. In ac measurements, the phase shift increases dramatically with frequency, especially for the high resistance case. We used five different ac frequencies from 11 Hz to 37 Hz for a mesoscopic two-terminal device made by wafer A. As shown in Fig. S7, there is no obvious difference between different frequencies, which indicates a rigorous ohmic behavior of the contacts.

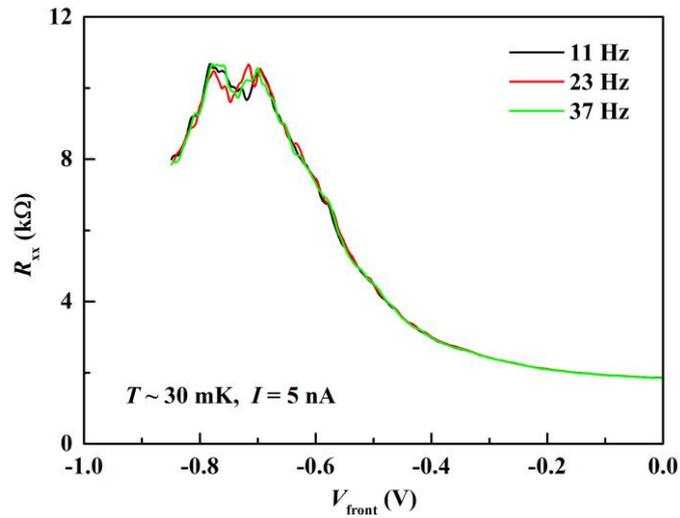

**Figure S7** $R_{xx}$-$V_{front}$ traces of a mesoscopic two-terminal device made by wafer A with different ac frequencies.